\documentclass[12pt,showpacs,showkeys,preprintnumbers,nofootinbib]{revtex4-1}

\usepackage{amssymb,amsmath,graphics,graphicx}

\begin{document}

\newcommand{\pderiv}[2]{\frac{\partial #1}{\partial #2}}
\newcommand{\deriv}[2]{\frac{d #1}{d #2}}

\title{Dynamics of epidemic spreading with vaccination: impact of social pressure and engagement}
 
\author{Marcelo A. Pires}
\thanks{pires$\_$ma@if.uff.br}

\author{Nuno Crokidakis}
\thanks{Corresponding author: nuno@if.uff.br}

\affiliation{
Instituto de F\'{\i}sica, Universidade Federal Fluminense, Niter\'oi - Rio de Janeiro, Brazil}

\date{\today}

\begin{abstract}
\noindent
In this work we consider a model of epidemic spreading coupled with an opinion dynamics in a fully-connected population. Regarding the opinion dynamics, the individuals may be in two distinct states, namely in favor or against a vaccination campaign. Individuals against the vaccination follow a standard SIS model, whereas the pro-vaccine individuals can also be in a third compartment, namely Vaccinated. In addition, the opinions change according to the majority-rule dynamics in groups with three individuals. We also consider that the vaccine can give permanent or temporary immunization to the individuals. By means of analytical calculations and computer simulations, we show that the opinion dynamics can drastically affect the disease propagation, and that the engagement of the pro-vaccine individuals can be crucial for stopping the epidemic spreading. The full numerical code for simulate the model is available from the authors' webpage.

\end{abstract}


\maketitle

\section{Introduction}

\qquad Epidemic spreading and opinion formation are two dynamical processes that have been atracted the interest of the scientific community in the last decades \cite{Barrat2008,Galam2012,Sen2013,Castellano2009,Satorras2015,JWu2015,NunoBJP,Nuno_pmco,Ni,zhou}. The interest of physicists varies from theoretical aspects like critical phenomena \cite{Moreno2002,Satorras2001,Silva2013,CrokidakisMenezes2012,Janssen2004,Biswas2011}, stochasticity \cite{Souza2010,Tomé2015}, universality \cite{Chung2015} and multiple phase transitions \cite{Mata2015}, to practical questions like the detection of the zero patient \cite{Antulov2015}, super spreaders \cite{Kitsak2010}, effects of self isolation \cite{CrokidakisQueirós2012} and others. More recently, the coupling of epidemic and opinion models have also been considered \cite{Funk2010,Wang2015,Perra2011,Manfredi}. 

Regarding a vaccination campaign in a given population, the individuals consider some points in order to make the decision to take the vaccine or not. In the case when a considerable fraction of the population decides to not take the vaccine, the consequences for the whole population may be drastic. As an example, in 2010 the French govenment requested vaccine for H1N1 for 90 million individuals, but about 6 million of the vaccines were effectively used by the population, and in this case the disease has spread fast \cite{Galam2010}. In this case, one can see that the public opinion can be a key feature in the diffusion of a disease in a given population, promoting the occurrence or the lack of an outbreak.

The public opinion about vaccination can be affected by economic factors. For example, we have a competition between the ``cost'' to become vaccinated (collateral effects, required time to take the vaccine, ...) and the injury caused by the disease when the individual did not take the vaccine (medication, money, miss some days of work, ...). In this case, the usual approach is to consider game theory or epidemiological-economic models \cite{Feng2011,Perrings2014}. However, usually the individuals/agents do not take into account only economic factors \cite{Funk2010, Wang2015, Perra2011, CoelhoCodeco2009, Han2014, Liu2012}. As discussed in \cite{Xia2013}, \textit{``if individuals are social followers, the resulting vaccination coverage would converge to a certain level, depending on individuals' initial level of vaccination willingness rather than the associated costs.''} In addition, in \cite{voinson} it is discussed that \textit{``assumptions of economic rationality and payoff maximization are not mandatory for predicting commonly observed dynamics of vaccination coverage such as the failure to reach herd immunity and oscillations between high and low levels of coverage''.} Related to social norms, the authors in \cite{oraby} propose that \textit{``including injunctive social norms will enable models of parental vaccinating behaviour for paediatric infectious diseases to better explain the whole range of observed vaccinating behaviour, including both vaccine refusal and the high vaccine coverage levels so commonly observed''.}

Indeed, some other works have shown that individuals are influenced by their social contacts in the process of opinion formation about a vaccination process \cite{Lau2010,Bish2011}. In this case, in this work we consider an opinion formation process coupled with an epidemic dynamics where vaccination is taking into account. Our target is to investigate how the density of Infected individuals in short and long times is affected by the social pressure and the engagement of the individuals regarding the vaccination. Thus, we are interested in answer some theoretical and practical questions:
\begin{itemize}
\item [(i)] What is the effect of social pressure and engagement in the epidemic spreading process?
\item [(ii)] What are the conditions for the occurrence of epidemic outbreaks in short times?
\item [(iii)] What is the critical initial density of pro-vaccine individuals that can avoid the occurrence of such short-time outbreaks?
\item [(iv)] The disease will survive in the long-time limit?
\end{itemize}
\noindent
The answer for these questions are given in the next sections.


\section{Model}

\qquad An individual's willingness to vaccinate is derived from his perception of disease risk and vaccine safety. However, the interactions among individuals in small groups will also affect the decision of the individuals to take or not the vaccine. In this case, we will consider an epidemic dynamics coupled with an opinion dynamics regarding the vaccination. Thus, we consider a fully-connected population with $N$ individuals or agents, that can be classified as follows:  

\begin{itemize}
\item Opinion states: Pro-vaccine (opinion $o=+1$) or Anti-vaccine (opinion $o=-1$) individuals;
\item Epidemic compartments: Susceptible (S), Infected (I) or Vaccinated (V) individuals;
\end{itemize}

Each opinion is supported by a given fraction of the population, namely $f_{+1}$ and $f_{-1}$, representing the fraction of Pro-vaccine and Anti-vaccine agents, respectively. We define the initial density of $+1$ opinions as $D$, that is a parameter of the model, and in this case the density of $-1$ opinions at the beginning is $1-D$. There are many models of opinion dynamics in literature \cite{JWu2015,NunoBJP,Nuno_pmco,Biswas2011,Galam2010,galam_mr,yang1,yang2,yu,ju,cao,qian,Galam1999}, and as a simple modeling of such dynamics we considered that the opinion changes are ruled by the majority-rule dynamics \cite{Galam1999,galam_mr}, i.e., we choose at random a group of 3 agents. If there is a local majority ($2\times 1$) in favor of one of the two possible opinions, the individual with minority opinion will follow the local majority. In this case, we are considering a mean-field formulation for the opinion dynamics. In the following we will see that the epidemic dynamics is also defined at a mean-field level. Despite the simplicity of a mean-field approach, it allows us an analytical treatment, that is important to a better understanding of a new model. Topologically, the mean-field approach corresponds to a fully-connected population, where each individual interacts with all others. In this case, it is also a realistic situation thanks to the modern communication networks \cite{naming}. Finally, it has been discussed that one can capture most of the dynamics of an epidemic on a real social network using only mean-field calculations \cite{bottcher}.

Regarding the epidemic dynamics, we made some assumptions. First of all, the opinion of an agent about the vaccination process determines his behavior regarding the decision to take the vaccine or not \cite{JWu2015,Ni,Funk2010,Wang2015,Feng2011,CoelhoCodeco2009}. As discussed in \cite{CoelhoCodeco2009}, \textit{``after conducting large scale studies on the acceptance of the Influenza vaccine, Chapman et al. \cite{chapman} conclude that perceived side-effects and effectiveness of vaccination are important factors in people's decision to vaccinate''.} We also considered that at the same time the disease is introduced in the population, a mass vaccination campaign is started. This is a realistic assumption, since usually the governements act fast in order to avoid disease outbreaks. For simplicity, we did not consider competition for doses. Finally, we considered as some authors \cite{voinson} that both dynamics (opinion and epidemic) occurs at the same time scale, i.e., the opinions evolve in the population due to interactions among agents, and at the same time a vaccination campaign occurs and the individuals may move among the epidemic compartments, that are defined in more details in the following. All these assumptions simplify the problem and makes the following analytical treatment easier. Furthermore, they are realistic, and were also considered by some authors from epidemiologists to mathematicians \cite{JWu2015,CoelhoCodeco2009,voinson}.

\begin{figure}[t]
\centering
\includegraphics[width=0.6\linewidth]{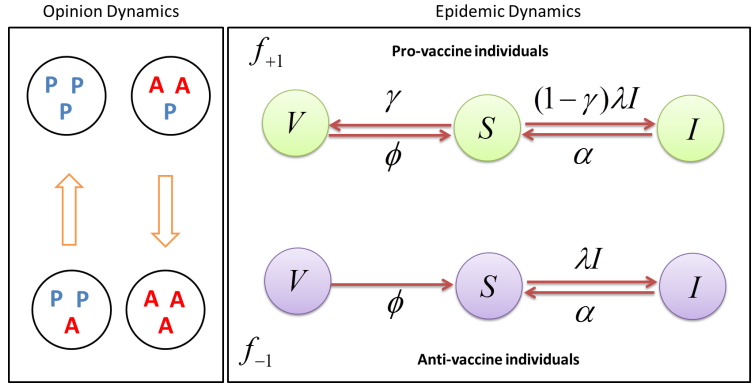}
\caption{Schematic representation of the model. It is shown the opinion dynamics of the Pro-vaccine (P) and Anti-vaccine (A) individuals, based on the majority rule (left panel), as well as the epidemic dynamics based on a compartmental model (right panel).}
\label{fig1}
\end{figure}

Now, let us elaborate upon the coupling of the two distinct dynamics (opinion and epidemic). Figure \ref{fig1} shows an esquematic representation of the dynamics. The Pro-vaccine agents (opinion $o=+1$) take the vaccine with probability $\gamma$. This parameter can be viewed as the engagement of the individuals regarding the vaccination campaign, i.e., it measures the tendency of an agent to go to the hospital to take a dose of the vaccine. Indeed, many times the individuals refuse to leave home to take a vaccine due to the time expended to conclude all the process. In this case, the complementary probability $1-\gamma$ will represent this ``laziness''. In the case a given individual does not take the vaccine, he can become infected with probability $\lambda$ if he make a contact with an Infected individual, as in a standard SIS model \cite{bailey,anderson}. In the same way, an Infected individual becomes Susceptible again with probability $\alpha$. Considering the Vaccinated agents, we considered that the vaccine is not permanent, so a vaccinated agent becomes susceptible again with rate $\phi$, the resusceptibility probability \cite{Shaw2010,Zanette2002}. Summarizing, the individuals with opinion $o=+1$ can undergo the following transitions among the epidemic compartments:
\begin{itemize}
\item $S \rightarrow V$: each Susceptible and Pro-vaccine individual becomes Vaccinated with probability $\gamma$;
\item $S \rightarrow I$: each Susceptible and Pro-vaccine individual becomes Infected with probability $(1-\gamma)\lambda$ if he is in contact with an Infected agent.
\end{itemize}

On the other hand, the agents against the vaccination process do not take the vaccine. In this case, they follow a standard SIS dynamics, with infection probability $\lambda$ and recovery probability $\alpha$ \cite{bailey,anderson}. Finally, the common transitions among states for both agents ($o=+1$ and $o=-1$) are given by 
\begin{itemize}
\item $I \rightarrow S$: each Infected individual recovers and becomes susceptible again with probability $\alpha$;
\item $V \rightarrow S$: each Vaccinated individual becomes Susceptible again with the resusceptibility probability $\phi$, since the vaccine wears off \cite{Shaw2010}.
\end{itemize}

As discussed above, as initial conditions we considered a fraction $D$ ($1-D$) of individuals carrying the opinion $o=+1$ ($o=-1$) at $t=0$. In addition, $1\%$ of the individuals start the dynamics in the Infected state, and the remaining $99\%$ in the Susceptible compartment. We considered synchronous update schemes for opinion and epidemic dynamics, i.e., both the transitions among the opinions and among the epidemic compartments are updated in a parallel way. For each simulation time step, the algorithm is as follows:
\begin{itemize}
\item We visit every site $i$ in sequential order, and for each site $i$ we choose at random other two sites $j$ and $k$. Then we apply the majority rule to each group of 3 agents ($i,j,k$).
\item the opinions are updated using a parallel scheme;
\item after the opinion dynamics, we visit every site in a sequential order;
\item we apply the epidemic dynamics' rules to each agent;
\item the epidemic states of the individuals are updated using a parallel scheme.
\end{itemize}

\vspace{0.3cm}

For the epidemic dynamics, the numerical procedure is as follows. For a spontaneous transition (with no contact, for example $S\rightarrow V$ or $I\rightarrow S$), the associated probability ($\gamma$ and $\alpha$ for the mentioned transitions, respectively) is compared with a uniformly distributed random number. In the case where a direct contact is needed, for example if we take a Susceptible agent, as we are considering a fully-connected network we choose at random another agent. If this agent is Infected, the transition $S\rightarrow I$ occurs with probability $\lambda$. For more details of the numerical procedure, see \cite{c_code}.

The mean-field equations for the model can be written as 
\begin{eqnarray} \label{eq1}
\frac{d\,S}{dt} & = & - \gamma\,S\,f_{+1} - (1-\gamma)\,\lambda\,S\,I\,f_{+1} - \lambda\,S\,I\,f_{-1} + \alpha\,I + \phi\,V ~, \\ \label{eq2}
\frac{d\,I}{dt} & = & (1-\gamma) \lambda S\,I\,f_{+1} + \lambda\,S\,I\,f_{-1} - \alpha\,I ~,\\ \label{eq3}
\frac{d\,V}{dt} & = & \gamma S\,f_{+1} - \phi\,V ~,
\end{eqnarray}
\noindent
where $S$, $I$ and $V$ are the fractions of Susceptibles, Infected and Vaccinated individuals, respectively. All the analytical calculations are performed in details in the Appendix \ref{app}. In the following section we discuss the main analytical results and the outcomes of our simulations.


\section{Results}

\qquad After the definition of the model, one can start to answer the four questions formulated at the end of the Introduction. As we are considering a fully-connected network, we considered the above mean-field equations to analyze the system, Eqs. (\ref{eq1}) - (\ref{eq3}), with special attention to the stationary properties of the model. We also considered an agent-based modelling of the system, since the individuals (agents) are the primary subject in a social theory \cite{Conte2012}. For this purpose, we considered populations with $N=10^4$ agents, and for sake of simplicity and without loss of generality \footnote{As it is usual in epidemic models, we verified that the effect of variation of $\alpha$ is small in the dynamics}., we fixed the recovery probability $\alpha=0.2$ in all simulations. Considering the stationary states of the model, all of our results are averaged over $100$ independent simulations in order to obtain better statistics. 

In the following we will consider separately the short-time and the long-time behavior of the system.


\subsection{Short-time behavior}

\qquad Let's start with the short-time behavior of the model, which allows us to analyze the question (i) made in the Introduction. As discussed in the previous section, our initial conditions are $I(0)=I_{o}=0.01$, $S(0)=S_{o}=0.99$, $V(0)=V_{o}=0.0$, $f_{+1}(0)=D$ and $f_{-1}(0)=1-D$. Thus, one can derive an analytical expression for the effective reproductive number $R_{e}$ through Eq. (\ref{eq2}), 

\begin{figure}[t]
\centering
\includegraphics[width=0.60\textwidth]{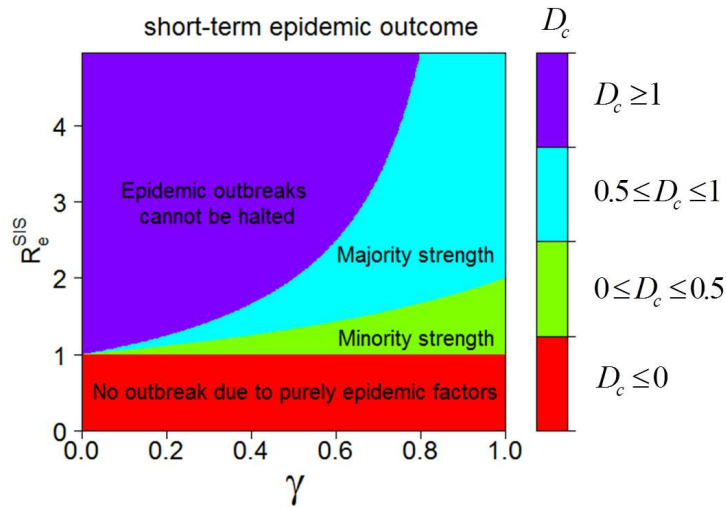}
\caption{Diagram of possible short-time epidemic scenarios as a function of the critical density $D_{c}$. We considered Eqs. (\ref{eq6}) and (\ref{eq7}) with parameters $S_{o}=0.99$, $I_{o}=0.01$, $\alpha=0.2$, and $\lambda$ was varied from $0$ to $1$. In this case, we obtained $R_e^{SIS}$ in the range $[0,4.95]$.}
\label{fig2}
\end{figure}

\begin{eqnarray} \label{eq4}
\frac{d\,I}{dt}\bigg|_{t=0} = (1-\gamma) \lambda\,S_{o}\,I_{o}\,f_{+1}(0) + \lambda S_{o}\,I_{o}\,f_{-1}(0) - \alpha\,I_{o} = I_{o}\,\alpha\,(R_{e} - 1) ~,
\end{eqnarray}
where $R_{e}$ is given by 
\begin{equation} \label{eq5}
R_{e}=(1-\gamma D)\,S_{o}\,\frac{\lambda}{\alpha} ~.
\end{equation}
Notice that the above expression depends on the initial fraction $D$ of Pro-vaccine agents and on the engagement $\gamma$, and it does not depend on the resusceptibility probability $\phi$. In other words, Eq. (\ref{eq5}) exhibits explicitly the coupling of the opinion and epidemic dynamics, including the usual dependence of the effective reproduction number $R_{e}$ on the initial condition $S_{o}$. If $S_{o}=1$ we have the basic reproduction ratio $R_{o}$ that is defined as the average number of individuals one infectious person would infect (over their entire period of infectiousness) if everyone in the population were susceptible \cite{keeling}.

Looking for Eq. (\ref{eq5}), one can see that the occurrence of an outbreak decays for increasing values of $\gamma$ and $D$. In this case, the initial fraction of vaccine supporters and the people engagement are important keys in the prevention of outbreaks, that is a realistic feature of the model. Observe that for $\gamma=0$ (no vaccination), the above expression (\ref{eq5}) recovers the usual SIS result, $R_{e}^{SIS}=S_{o}\,\lambda/\alpha$.

In order to compare both expression for the effective reproductive number, one can rewrite Eq. (\ref{eq5}) as
\begin{equation} \label{eq6}
R_{e} = (1-\gamma D)\,R_{e}^{SIS} ~.
\end{equation}

To avoid an epidemic outbreak, $R_{e}\leq 1$, which implies that $D\geq(1/\gamma)(1-1/R_{e}^{SIS})$. In this case, one can define a critical density $D_{c}$ as
\begin{equation} \label{eq7}
D_{c}=\frac{1}{\gamma}\,\left(1-\frac{1}{R_{e}^{SIS}}\right)  ~.
\end{equation}
\noindent
Figure \ref{fig2} exhibits a diagram that help us to answer the question (ii) pointed in the Introduction, as discussed in the following:
\begin{enumerate}
\item Red region ($D_{c} \leq 0$): in this region the solution (\ref{eq7}) is mathematically valid but it presents no physical meaning. It indicates that the outbreak does not occur due to purely epidemic reasons, i.e., we have $R_{e}^{SIS}<1$ since $S_{o}\,\lambda<\alpha$;

\item Green region ($0<D_{c} \leq 0.5$): in this region the initial minority is favorable to the vaccination. However, this minority can avoid the epidemic outbreak, and one can see the \textit{power of the initial minority in the short time} \cite{Nuno_pmco,Huang2008,Huang2009}.

\item Blue region ($0.5 \leq D_{c} < 1$): in this region, the initial majority is in favor of the vaccination. Thus, depending on the engagement of such majority the outbreak can be avoided, and one can see the \textit{power of the initial majority in the short time} \cite{Huang2008,Huang2009}.

\item Purple region ($D_{c}\geq 1$): in this region the solution (\ref{eq7}) is also not physically acceptable. It indicates that the outbreak can not be avoided, even if $100\%$ of the population is in favor of the vaccination. Values $D_{c}\geq 1$ are obtained if the people engagement is too low.
\end{enumerate}
\noindent
Those last results answer the questions (ii) and (iii) made in the Introduction.


\subsection{Long-time behavior}

\qquad In this section we analyze the long-time behavior of the model, i.e., its steady-state properties, and one can discuss the question (iv) pointed in the Introduction. As we will see in the following, in opposition to what happend in the short-time case, the stationary behavior of the model depends on the resusceptibility probability $\phi$. In this case, we will study separately the two cases $\phi\neq 0$ and $\phi=0$. 


\subsubsection{Vaccination with limited efficiency ($\phi \neq 0$)}

\quad For the case with temporary immunity $\phi \neq 0$, one can obtain all the fractions of the epidemic states in the limit $t\to\infty$ (see Appendix \ref{app}). The stationary density of Vaccinated and Susceptible individuals are given, respectively, by
\begin{align}
V_{\infty}  =
\begin{cases} 
0 & if \ D<0.5
\\
 \frac{\gamma}{\phi}\,\frac{\alpha }{(1-\gamma)\,\lambda}
 & if \ D>0.5
\end{cases}
\label{eq8}
\end{align} 

\begin{align}
S_\infty  =
\begin{cases}
\frac{\alpha}{\lambda} & if \  D<0.5 
\\ \frac{\alpha}{\lambda\,(1-\gamma)}  &  if \ D>0.5
\end{cases}
\label{eq9}
\end{align} 

On the other hand, the stationary density of Infected individuals is given by
\begin{align} \label{eq10}
I_\infty  =
\begin{cases} 0 & if \ \lambda \leq \lambda^{I}_c \ and \ D<0.5        \ or \       
\ \lambda \leq \lambda^{II}_{c} \ and \ D>0.5
\\  
\frac{1}{\lambda}\,(\lambda-\lambda^{I}_{c}) & 
if \ \lambda > \lambda^{I}_{c} \ and \ D<0.5
\\ 
\frac{1}{\lambda}\,(\lambda-\lambda^{II}_{c}) & 
if \ \lambda > \lambda^{II}_{c}  \ and \ D>0.5
\end{cases}
\end{align} 
where the epidemic thresholds for the cases $D<0.5$ and $D>0.5$ are, respectively,
\begin{eqnarray} \label{eq11}
\lambda^{I}_{c} & = & \alpha ~, \\ \label{eq12}
\lambda^{II}_{c} & = & \alpha\,\frac{\phi + \gamma}{\phi\,(1-\gamma)} ~.
\end{eqnarray}
\noindent
Equations (\ref{eq11}) and (\ref{eq12}) show that effect of the social pressure is twofold. For $D<0.5$, the social pressure has a negative effect, in a way that it eliminates from the threshold $\lambda^{I}_{c}$ the effect of the engagement $\gamma$ of the Pro-vaccine individuals. On the other hand, for $D>0.5$ the social pressure has a positive effect, since the contribution of the engagement $\gamma$ appears explicitly in the epidemic threshold $\lambda^{II}_{c}$. The effect of the engagement can be analyzed in more details as follows. Taking $\lambda^{II}_{c}=1$ in Eq. (\ref{eq12}) we found a threshold value
\begin{equation} \label{eq13}
\gamma^{*} = \frac{\phi\,(1-\alpha)}{\phi+\alpha} ~.
\end{equation}
\noindent
This value $\gamma^{*}$ is the engagement above which there is no endemic phase in the system.

\begin{figure}[t]
\centering
\includegraphics[width=0.36\textwidth]{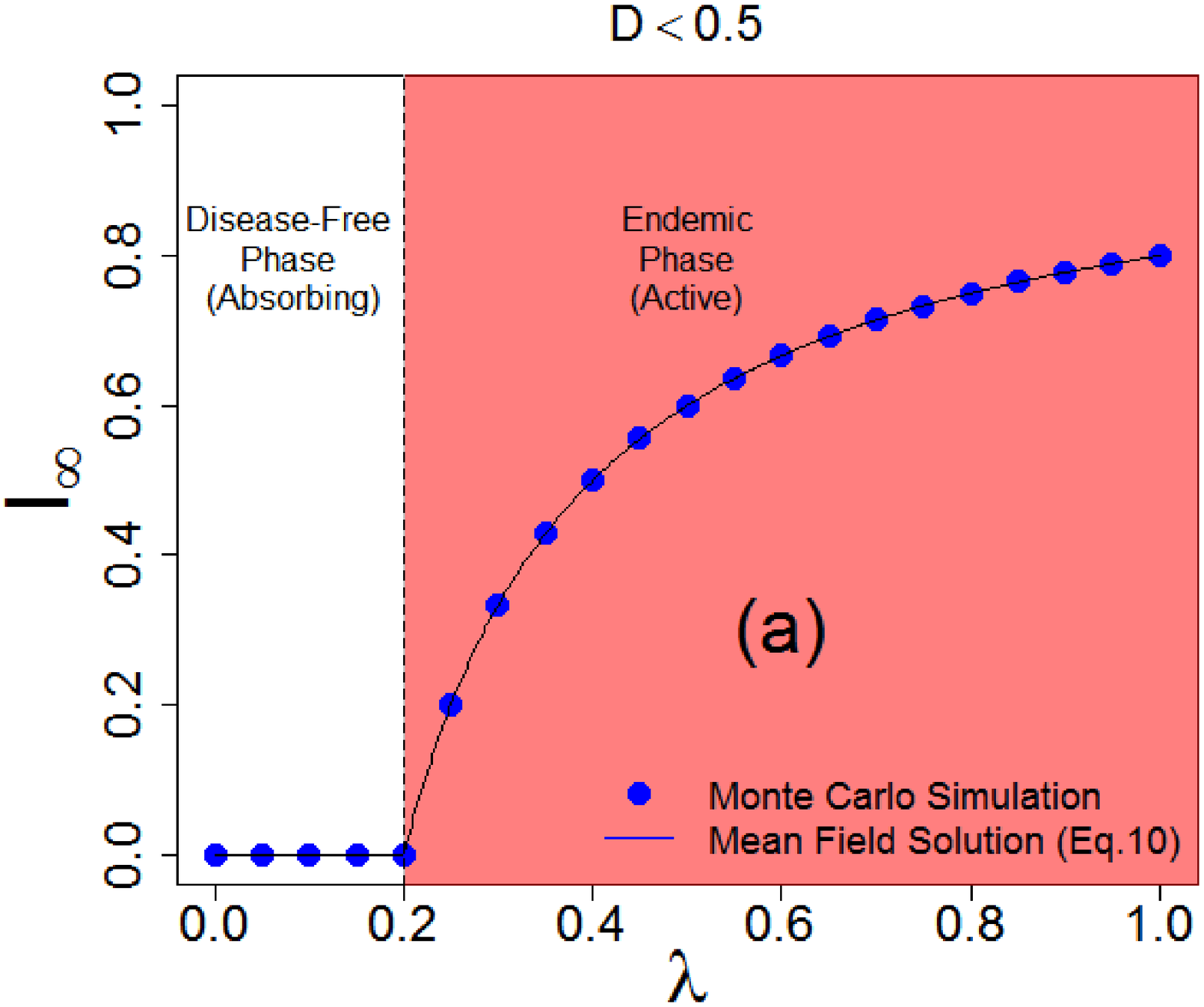}
\includegraphics[width=0.36\textwidth]{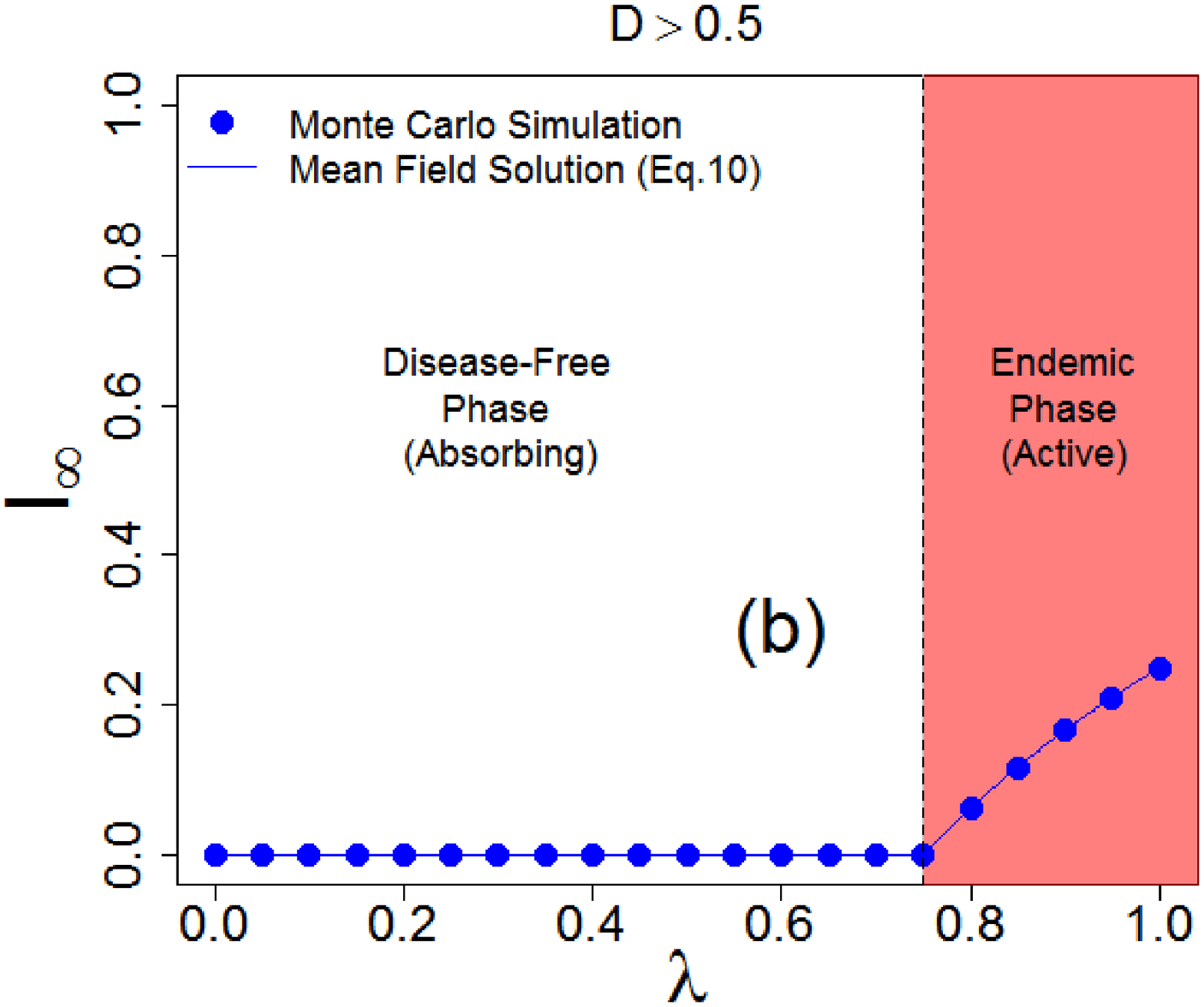}
\\
\includegraphics[width=0.36\textwidth]{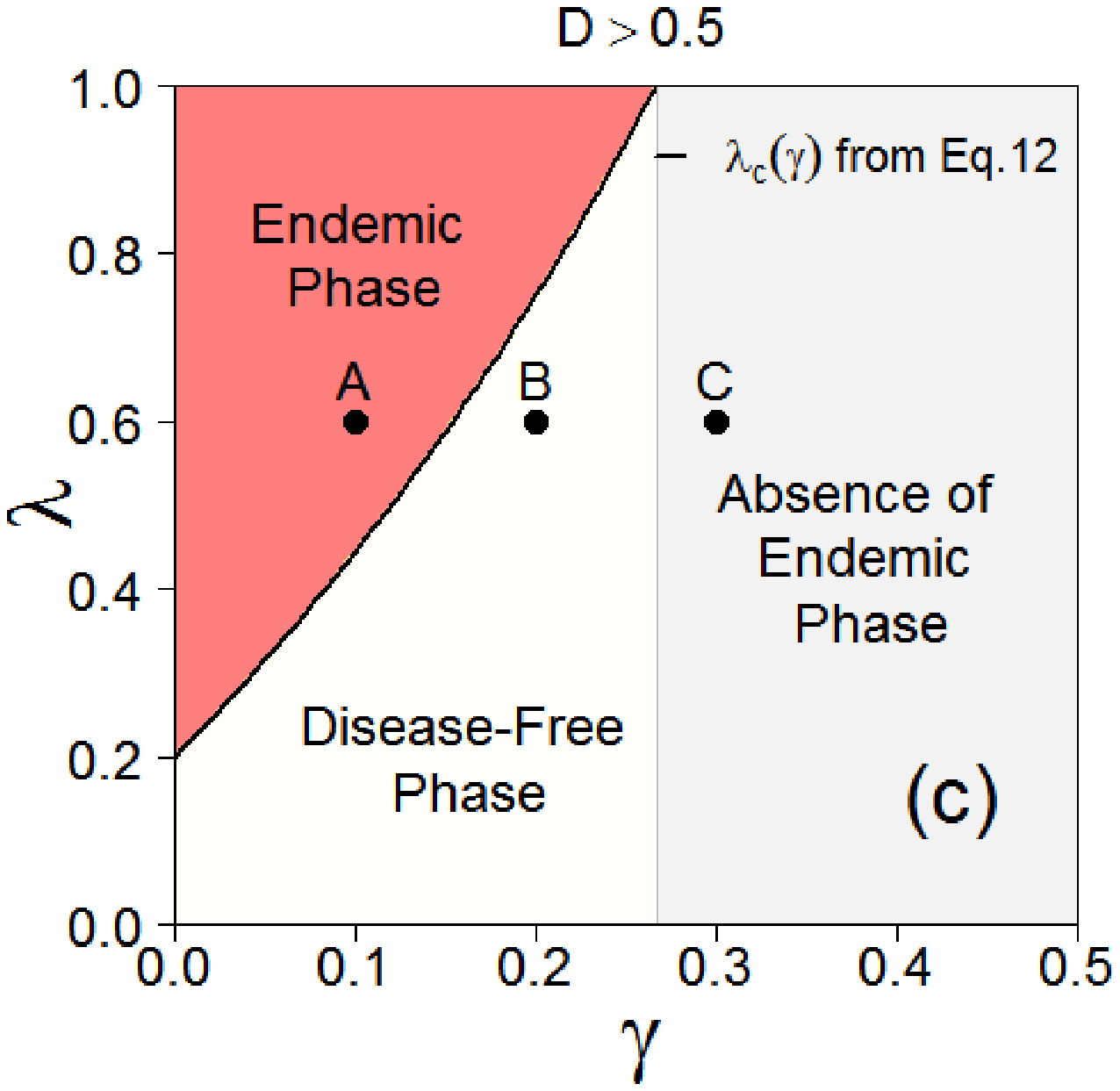}
\includegraphics[width=0.36\textwidth]{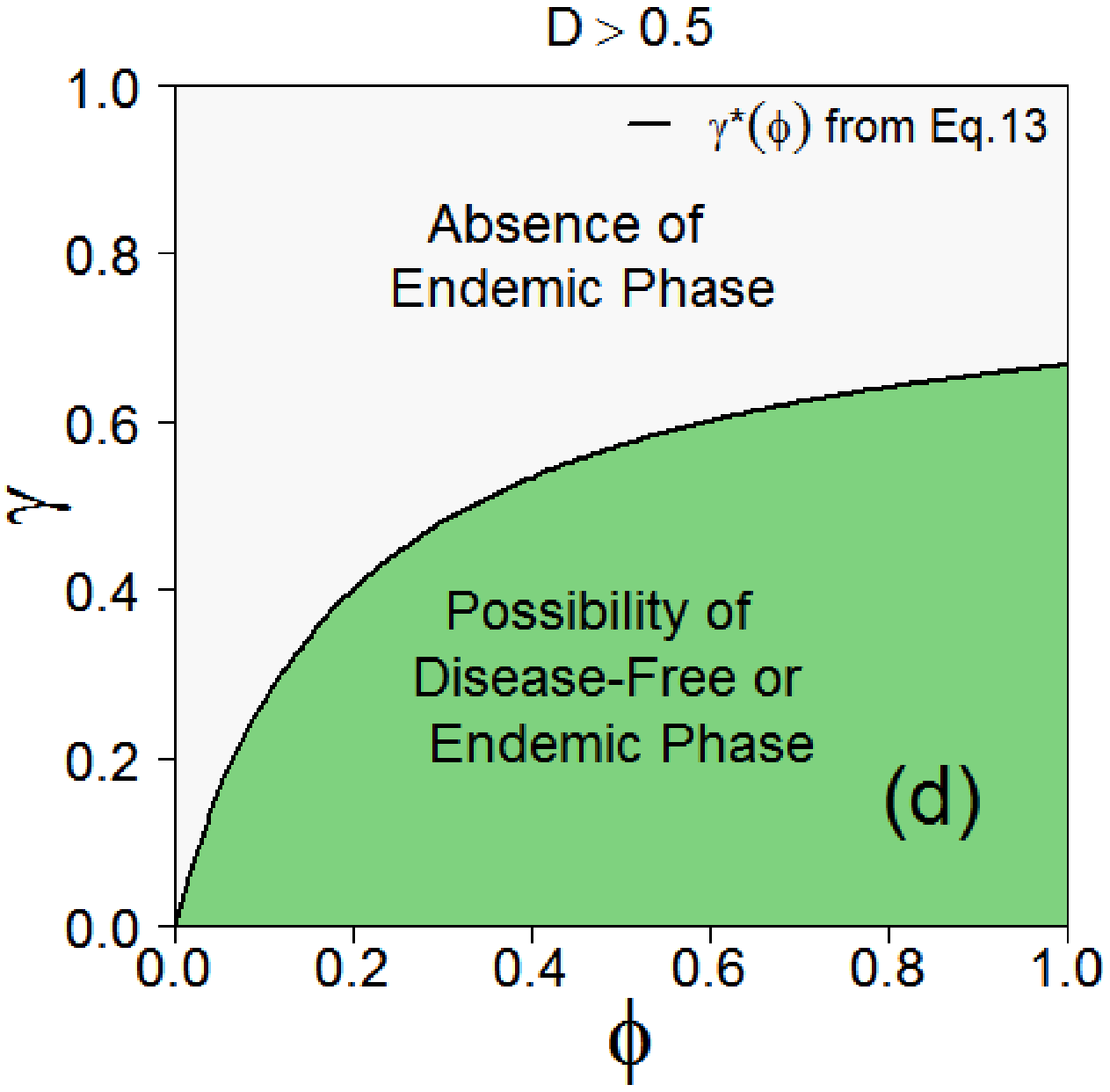}
\caption{Typical shape of the stationary density $I_{\infty}$ of Infected individuals as a function of $\lambda$ for $D<0.5$ {\bf (a)} and $D>0.5$ {\bf (b)}. We considered $\alpha=0.2$, $\phi=0.1$ and $\gamma=0.2$. The line is the analytical result Eq. (\ref{eq10}) and the points were obtained from simulations using $D=0.4$ {\bf (a)} and $D=0.6$ {\bf (b)}. {\bf (c)} Typical phase diagram $\lambda$ versus $\gamma$ for $D>0.5$ given by Eq. (\ref{eq12}). The parameters are $\phi=0.1$ and $\alpha=0.2$. {\bf (d)} Typical phase diagram $\gamma$ versus $\phi$ for $D>0.5$ given by Eq. (\ref{eq13}). The parameter is $\alpha=0.2$.}
\label{fig3}
\end{figure}

In Figure \ref{fig3} we exhibit some analytical and numerical results. In Figures \ref{fig3} (a) and (b) we show typical results for the stationary density of Infected individuals $I_{\infty}$ for $D<0.5$ and $D>0.5$, respectively. The full lines are given by the solutions (\ref{eq10}), with the respective epidemic thresholds given by Eqs. (\ref{eq11}) and (\ref{eq12}), and the symbols are results of numerical simulations of the model. In those figures one can see the phase transition between a Disease-free phase, where the disease disappears of the population after a long time, and an Endemic phase, where the disease survives and infects permanently a finite fraction of the population in the stationary state. According to Eq. (\ref{eq10}), in comparison with the standard form $I_{\infty}\sim(\lambda-\lambda_{c})^{\beta}$, we obtain a mean-field exponent $\beta=1$, as expected due to the mean-field character of our model. In other words, a typical active-absorbing nonequilibrium phase transition.

The critical points $\lambda_{c}$ do not depend explicitly on the initial density of Pro-vaccine agents, see Eqs. (\ref{eq11}) and (\ref{eq12}). However, for better understanding of the dynamics, it is important to analyze what is the initial minority opinion. For $D<0.5$, after a long time the Pro-vaccine agents (opinion $o=+1$) will disappear of the population, so their engagement is not relevant in this case. On the other hand, for $D>0.5$ the opposite occurs, in the steady state all agents will share the opinion $o=+1$. Thus, during the evolution of the dynamics, when the fraction of Pro-vaccine agents is increasing in time, an increase in the engagement probability $\gamma$ contributes to the extinction of the disease, or in other words to the increase of the Disease-free phase. This behavior can be observed in Eq. (\ref{eq12}): when we increase $\gamma$ the critical point $\lambda_{c}^{II}$ increases, and the endemic (disease-free) phase becomes smaller (larger). This picture can be seen in Figure \ref{fig3} (c), where we plot Eq. (\ref{eq12}) as a function of $\gamma$. Let us consider the points A, B and C in this figure. If there is an effort (public policies, for example) to increase the engagement of the people in favor of vaccination from $\gamma=0.1$ (point A) to $\gamma=0.2$ (point B), the disease will disappear after a long time. In addition, if the engagement rises again, for $\gamma=0.3$ (point C), there is no Endemic phase anymore, independent of the infection rate $\lambda$, since in this region of the figure we have $\lambda_{c}>1$. These results indicate that even for highly contagious diseases the Pro-vaccine movement can stops the epidemic spreading, provided that the related agents are sufficiently engaged.

In this case, other question immediately arises: the above-mentioned absence of an Endemic phase is robust under variations in the resusceptibility probability $\phi$? The graphic in Figure \ref{fig3} (d), where it is plotted Eq. (\ref{eq13}), answers this question. Indeed, for a sufficient high engagement one obtains the absence of the Endemic phase, independent of the vaccine's imperfection $\phi$. Nevertheless, one can also see in Figure \ref{fig3} (d) that for increasing values of $\phi$ it is necessary more engagement in order to eliminate the disease propagation, which is also a realistic result of the model.

\begin{figure}[t]
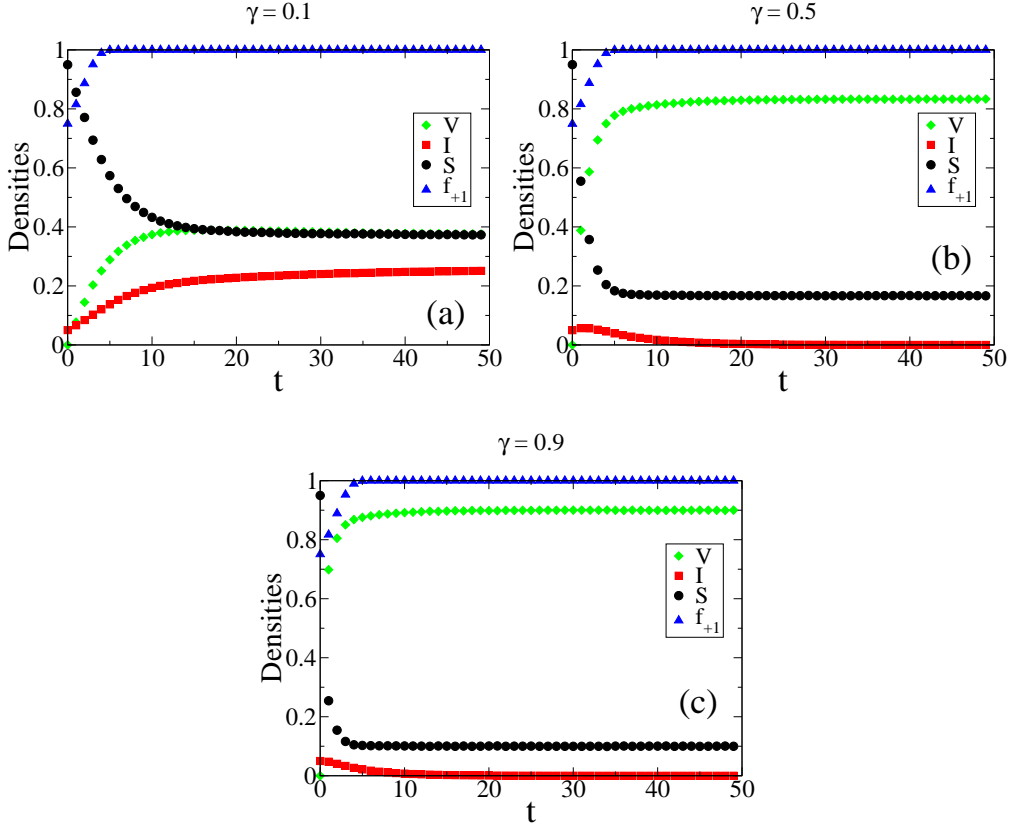

\centering
\includegraphics[width=0.4\textwidth]{figure4a.eps}
\includegraphics[width=0.4\textwidth]{figure4b.eps}
\\
\vspace{0.5cm}
\includegraphics[width=0.4\textwidth]{figure4c.eps}
\caption{Temporal evolution of the densities of Pro-vaccine ($f_{+1}$), Susceptible (S), Infected (I) and Vaccinated (V) individuals, obtained from numerical simulations of the model for $\gamma=0.1$ (a), $\gamma=0.5$ (b) and $\gamma=0.9$ (c). The other parameters are $\lambda=0.6$, $\phi=0.1$, $D=0.75$, $N=10^4$, $I_{o}=0.05$, $S_{o}=0.95$, $V_{o}=0$. From Eqs. (\ref{eq5}) and (\ref{eq12}), we obtain (a) $R_e=2.64$ and $\lambda_{c}=0.44$, (b) $R_e= 1.78$ and $\lambda_{c}=0.75$, (c) $R_e= 0.93$ and $\lambda_{c}=20.0$.}
\label{fig4}
\end{figure}

In Figure \ref{fig4} we exhibit numerical results for $D=0.75$, i.e., for an initial majority of agents supporting the vaccination process. In this case, the opinion dynamics leads the system to a steady state were all agents follow this initial majority. Even in this case, a small engagement of the individuals like $\gamma=0.1$ is not sufficient to avoid an epidemic outbreak ($R_e=2.64>1$ and $\lambda=0.6 > \lambda_c=0.44$), and in addition the disease survives in the population in the long-time limit. On the other hand, for $\gamma=0.5$ we have a disease-free equilibrium after a long time, but the outbreak still occurs ($R_e=1.78 >1$ and $\lambda=0.6 < \lambda_c=0.75$). Finally, for a sufficient high engagement like $\gamma=0.9$, the outbreak does not occurs ($R_e=0.93<1$ and $\lambda=0.6 < \lambda_c=20.0$), the epidemic disappears in few time steps of the population, and we have a disease-free equilibrium in the long-time limit.


\subsubsection{Vaccination with permanent immunization ($\phi=0$)}

\quad For the case with permanent immunity $\phi=0$, one can obtain only some analytical results for the fractions of the epidemic states, in the limit $t\to\infty$ (see Appendix \ref{app}). The stationary density of Vaccinated, Susceptible and Infected individuals are given, respectively, by

\begin{figure}[t]
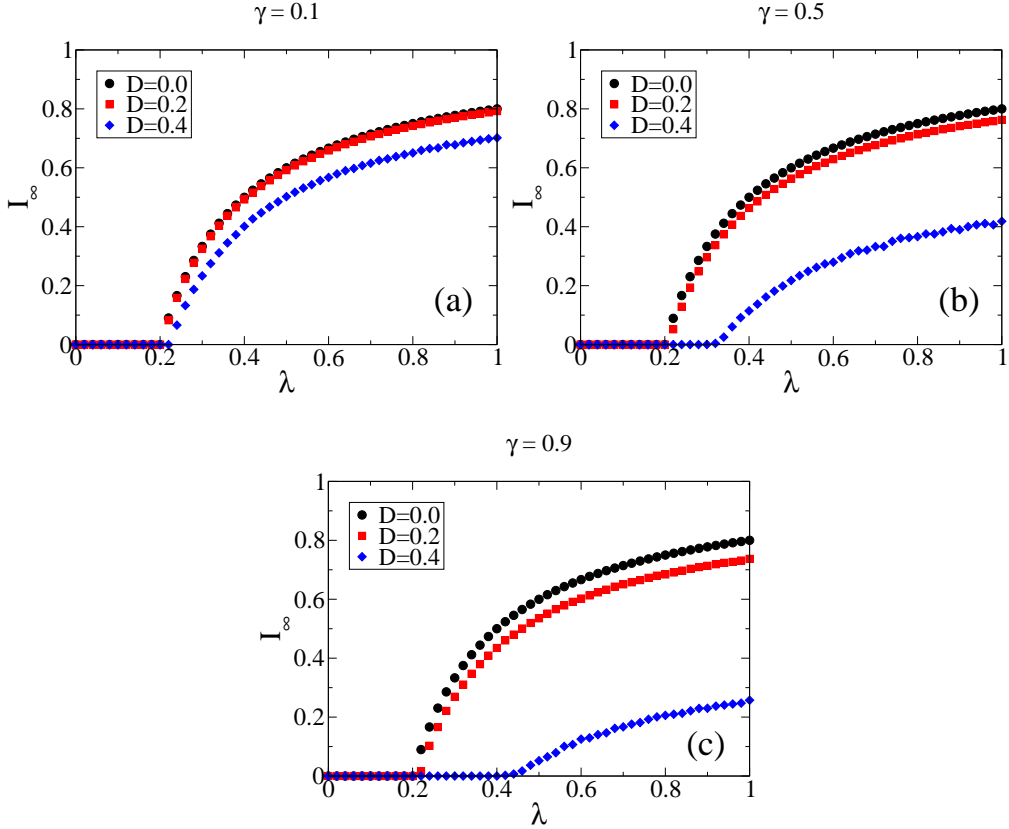

\centering
\includegraphics[width=0.4\textwidth]{figure5a.eps}
\includegraphics[width=0.4\textwidth]{figure5b.eps}
\\
\vspace{0.5cm}
\includegraphics[width=0.4\textwidth]{figure5c.eps}
\caption{Stationary density $I_{\infty}$ of Infected individuals as a function of $\lambda$ for typical values of $D=0.0$, $0.2$ and $0.4$. The numerical results where obtained for $\phi=0$ (permanent immunization) and $\gamma=0.1$ (a), $\gamma=0.5$ (b) and $\gamma=0.9$ (c). The population size is $N=10^4$ and data are averaged over $100$ independent simulations.}
\label{fig5}
\end{figure}

\begin{align}  \label{eq14}
V_\infty  =
\begin{cases} 
 V_\infty & if  \ D<0.5
\\
1
 &
if \  \ D>0.5
\end{cases}
\end{align} 

\begin{align}  \label{eq15}
S_\infty  =
\begin{cases} 
\frac{\alpha}{\lambda} &  if \  D<0.5  
\\ 
0 & if \  D>0.5  
\end{cases}
\end{align} 
\begin{align}  \label{eq16}
I_\infty  =
\begin{cases} 0 & if  \ D>0.5        \ or \    D<0.5 \ and \  (1-V_{\infty})\lambda \leq \alpha 
\\
1 - V_{\infty} - \frac{\alpha}{\lambda }  & if \ D<0.5  \ and \ (1-V_{\infty})\lambda \geq  \alpha 
\end{cases}
\end{align}

As one can see, for $\phi=0$ the above equations furnish only the exact solution for $S_{\infty}$, and a relation between $I_{\infty}$ and $V_{\infty}$  This is a consequence of the compartment $V$, that for $\phi=0$ is an absorbing state, i.e., if an agent enters in this compartment he will not change his epidemic state anymore. In this case, we will focus our study mainly in computer simulations. However, the above analytical results can give us some insights in the behavior of the model. First of all, Eq. (\ref{eq16}) predicts a phase transition. On the other hand, one can see from Eq. (\ref{eq14})-(\ref{eq16}) that if the initial majority is formed by Pro-vaccine individuals ($D>0.5$), the disease will disapear after a long time, since the vaccine induces a permanent immunization and more agents become in favor of the vaccination during the evolution of the system. 

In Figure \ref{fig5} we show numerical results for the stationary density $I_{\infty}$ of Infected individuals as a function of $\lambda$ for typical values of $\gamma$ and $D$, with $\phi=0$. One can see that the epidemic threshold $\lambda_{c}$ increases for increasing values of $D$. In addition, one can see that when we increase $D$ from $D=0.0$ to $D=0.2$, the impact on the behavior of $I_{\infty}$ is small, whereas the effect is more pronounced when we change from $D=0.2$ to $D=0.4$. This effect is more clear when we increase the engagement $\gamma$, see for example Figure \ref{fig5} (c).

\begin{figure}[t]
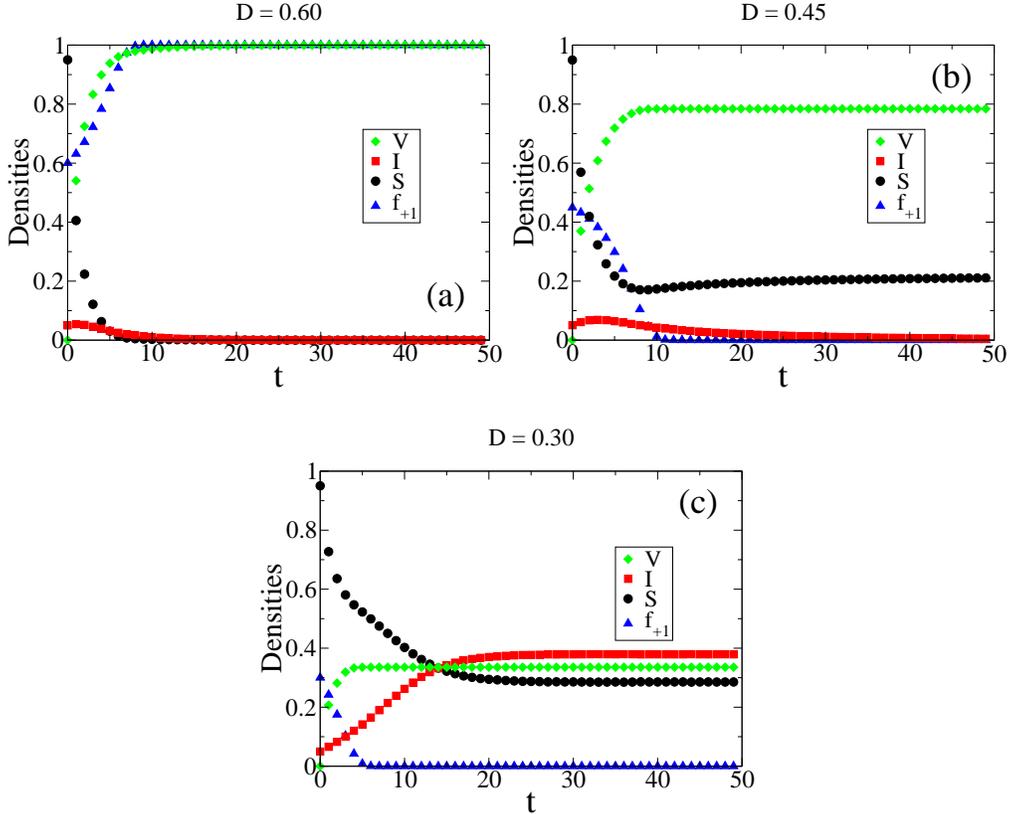

\centering
\includegraphics[width=0.4\textwidth]{figure6a.eps}
\includegraphics[width=0.4\textwidth]{figure6b.eps}
\\
\vspace{0.5cm}
\includegraphics[width=0.4\textwidth]{figure6c.eps}
\caption{Temporal evolution of the densities of Pro-vaccine ($f_{+1}$), Susceptible (S), Infected (I) and Vaccinated (V) individuals, obtained from numerical simulations of the model with permanent immunization ($\phi=0$). Results are for $D=0.6$ (a), $D=0.45$ (b) and $D=0.3$ (c). The other parameters are $\lambda=0.7$, $\gamma=0.9$, $N=10^4$, $I_{o}=5\%$, $S_{o}=95\%$, $V_{o}=0\%$. From Eq. (\ref{eq6}) one obtains $R_e=1.53$ (a), $R_e= 1.98$ (b) and $R_e= 2.43$ (c).}
\label{fig6}
\end{figure}

We exhibit in Figure \ref{fig6} numerical results for the temporal evolution of the densities of Pro-vaccine ($f_{+1}$), Susceptible (S), Infected (I) and Vaccinated (V) individuals for $\phi=0$. Results are for $D=0.6$ (a), $D=0.45$ (b) and $D=0.3$ (c), and the fixed parameters are $\lambda=0.7$ and $\gamma=0.9$. In the panel (a) one can see that even if the vaccination supporters are initially a majority ($D=0.6$), they cannot avoid the occurrence of the outbreak (since $R_{e}>1$), but they can promote the disappearance of the disease in the long-time limit. Nevertheless, due to the initial majority in favor of vaccination, the outbreak achieves a small fraction of the individuals. In Figure \ref{fig6} (b) there is a small initial majority against the vaccination ($D=0.45$). Despite the occurrence of an epidemic outbreak (since $R_{e}>1$), one can see an interesting result: the disease disappears in the long-time limit, even in this case ($D<0.5$) where after a long time all the individuals in the population will share the anti-vaccine opinion $o=-1$. This result occurs due to the high engagement of the population, $\gamma=0.9$, i.e., even being a minority in all the evolution of the system the Pro-vaccine individuals take the vaccine with a high probability, and the immunity is permanent. In this case, even when these individuals change opinion due to the social pressure, they are already vaccinated.

Finally, in the panel (c) of Figure \ref{fig6} we show results for $D=0.3$, i.e., the great majority of the population ($70\%$) starts the dynamics against the vaccination process. In this case, one can see that the outbreak occurs (since $R_{e}>1$) and the disease survives in the stationary state. This occurs due to the rapid population consensus against the vaccination, and even for a high engagement the pro-vaccine agents disappear rapidly of the population due to social pressure.

\begin{figure}[t]
\begin{center}
\vspace{0.5cm}
\includegraphics[width=0.4\textwidth,angle=0]{figure7a.eps}
\hspace{0.2cm}
\includegraphics[width=0.4\textwidth,angle=0]{figure7b.eps}
\\
\includegraphics[width=0.38\textwidth,angle=-90]{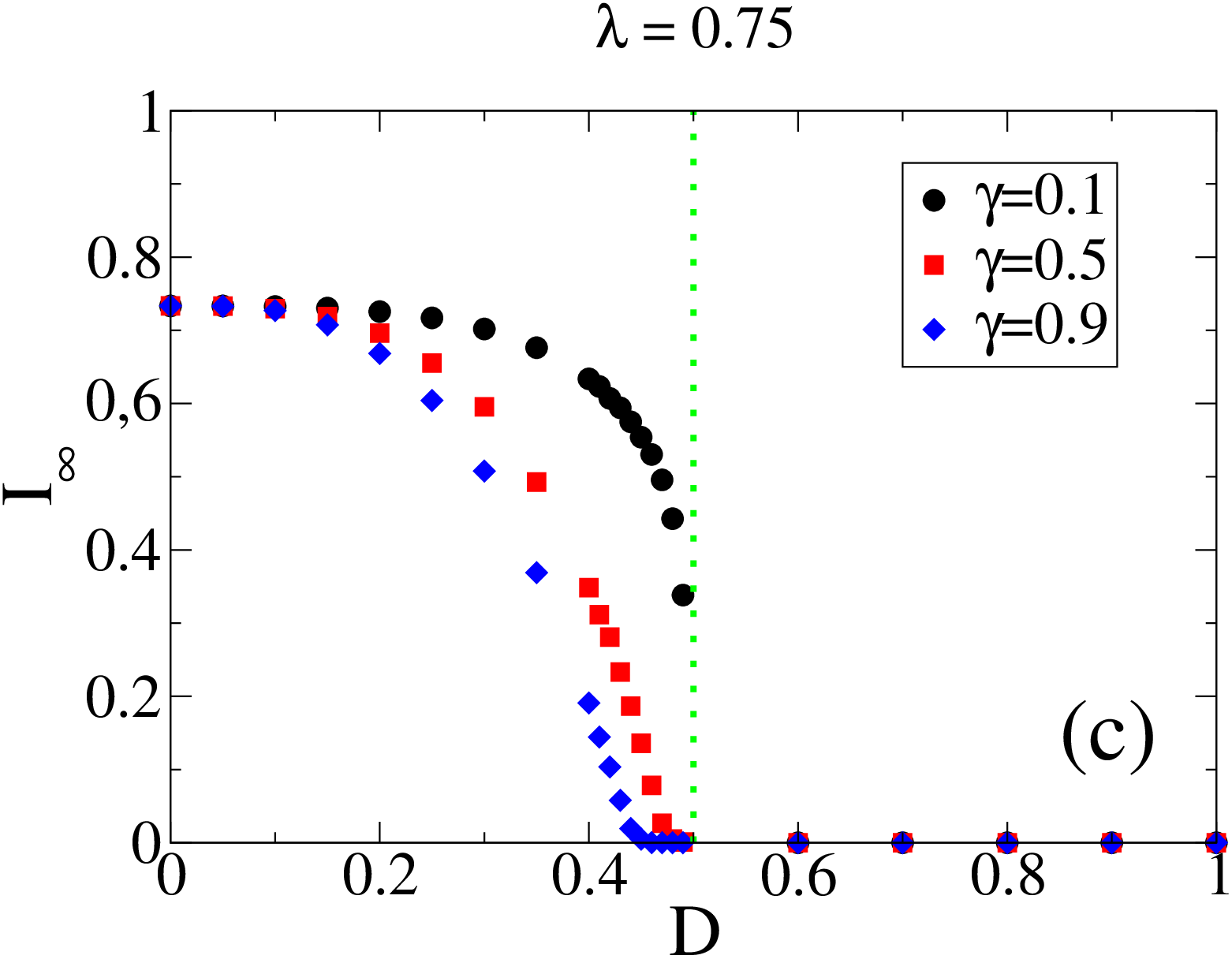}
\end{center}
\caption{Stationary density $I_{\infty}$ of Infected individuals as a function of the initial density $D$ of pro-vaccine agents for typical values of the engagement $\gamma$ and permanent immunity ($\phi=0$). Results are for infection probabilities $\lambda=0.25$ (a), $\lambda=0.50$ (b) and $\lambda=0.75$ (c). The vertical (green) dotted line indicates the value $D=0.5$, that separates the regions of initial majorities in favor and against the vaccination process. The population size is $N=10^{4}$ and data are averaged over $100$ independent simulations.}
\label{fig7}
\end{figure}

For better analyze the effect of social pressure in the epidemic spreading, we exhibit in Figure \ref{fig7} the stationary density $I_{\infty}$ of Infected individuals as a function of the initial density $D$ of $+1$ opinions, for typical values of infection probability $\lambda$ and of the engagement $\gamma$. First of all, one can see that, if the initial majority is in favor of the vaccination campaign ($D>0.5$), the disease disappears of the population in the steady state, independent of the values of $\gamma$ and $\lambda$, as predicted analytically in Eqs. (\ref{eq14})-(\ref{eq16}). This is a positive effect of the social pressure \cite{oraby,Wu2013,Genki2016}. On the other hand, for small values of $D$ ($D<\approx 0.15$) the density of Infected agents is independent of $\gamma$, for a fixed value of $\lambda$. In other words, if the the population is initially mainly dominated by individuals against the vaccination, the engagement has no effect on the disease propagation, and we have an endemic phase. In this case, the social pressure causes a negative effect on the disease spreading \cite{oraby,Wu2013,Genki2016}. However, one can see in Figure \ref{fig7} an interesting result for the cases with $D<0.5$: despite the mentioned negative effect of social pressure, a sufficient high engagement like $\gamma=0.9$ can lead to the disease extinction in the stationary state, since the initial majority of anti-vaccine individuals is not much larger than the initial minority of pro-vaccince agents. This result can be observed in the three graphics of Figure \ref{fig7}, and is more pronounced for the smaller values of the infection probability $\lambda$. These results can be seen as a kind of power of the minority in the long-time limit. All those results are realistic features of our model.


\section{Discussion}

\qquad In this work, we have studied a model of epidemic spreading coupled with an opinion dynamics. Such opinion dynamics simulates the competition of two distinct positions regarding a vaccination campaign: pro-vaccine x anti-vaccine individuals. As a minimal model, we considered that this competition is represented in our case by the Galam's majority-rule dynamics \cite{Galam1999,galam_mr}. In addition, the epidemic dynamics is governed by a compartmental model with 3 classes: Susceptibles (S), Infected (I) and Vaccinated (V) individuals. Finally, we considered two distinct cases: the vaccine can give temporary or permanent immunity to the individuals.

For both cases, one can describe the model at mean-field level through ordinary differential equations. First, we analyzed the short-time behavior of the model, and we found that it does not depend on the resusceptibility rate $\phi$, the probability for which a vaccinated individual becomes susceptible again. We calculated the effective reproductive number $R_{e}$, that is given by $R_{e}=(1-\gamma D)\,S_{o}\,(\lambda/\alpha)$. This expression shows that the occurrence or absence of an epidemic outbreak depends on the infection and recovery probabilities ($\lambda$ and $\alpha$, respectively), as usual, but it also depends on social parameters. The first one is the initial fraction $D$ of pro-vaccine individuals, and the other one is the engagement $\gamma$ of those individuals. Thus, the coupling of both dynamics (opinion and epidemic) affects directly the disease diffusion in the short time. Based on the expression for $R_{e}$, we discussed the range of values of $D$ for which different realistic scenarios can occur. In this case, we found evidences that even an initial minority in favor of the vaccination campaign ($D<0.5$) can stop the disease spreading, since its engagement is sufficiently high (the power of the initial minority). In addition, for the case where the pro-vaccine individuals are majority in the beginning ($D>0.5$), even the outbreak can be avoided, depending on the values of $\gamma$ (the power of the initial majority).

On the other hand, the long-time behavior of the model depends on $\phi$. For the cases where the immunity is temporary, we found all the stationary densities (S, I and V). Based on these results, we verified that the model undergoes phase transitions between disease-free and endemic phases, and the epidemic thresholds are different for the two cases of initial majorities ($D<0.5$ and $D>0.5$). Thus, we found that for $D<0.5$ the transition does not depends on $\phi$, which suggests that if the initial majority is against the vaccination process, the engagement does not influence the disease spreading. On the other hand, for $D>0.5$ the threshold depends explicity on $\phi$, and we found that for sufficient high engagement the disease can disappear of the population in the long time, even for high infection rates. We also derived an analytical expression for the threshold value $\gamma^{*}$ above which there is no endemic phase in the population.

For the cases where the immunity is permanent, we only derived a closed analytical expression for the stationary density of Susceptible agents $S_{\infty}$. The expression for $I_{\infty}$ depends on $V_{\infty}$, and we can conclude from that expressions only that there is a phase transition as in the previous case, and that if the initial majority is formed by Pro-vaccine individuals ($D>0.5$), the disease will disapear after a long time, since the vaccine induces a permanent immunization and more agents become in favor of the vaccination during the evolution of the system due to the social pressure (majority-rule dynamics). This is a positive effect of the social pressure. The remaining of the analysis for permanent immunization was done through Monte Carlo simulations. In this case, we also found distinct epidemic scenarios, with endemic and disease-free phases, and the occurrence or not of outbreaks. In addition, we verified that if there is initially an overwhelming majority of anti-vaccine agents ($D<<0.5$), the engagement has no effect on the disease propagation, and we have an endemic phase, which can be interpreted as a negative effect of the social pressure on the disease spreading. However, even for $D<0.5$ a sufficient high engagement can lead to the disease extinction in the stationary state, since the initial majority of anti-vaccine individuals is not much larger than the initial minority of pro-vaccince agents. These results can be seen as a kind of power of the minority in the long-time limit. All these those results are realistic features of our model.

As discussed in \cite{Zanette2002} the variation of the epidemic threshold as a result of immunization has important practical consequences. In our case, for $\phi\neq 0$ (temporary immunity), one can see in Figure \ref{fig3} (c) that near the engagement threshold $\gamma^{*}$ a small effort to increase the engagement of pro-vaccine individuals leads the system to a disease-free phase, or even to the absence of an endemic phase (i.e., there is no endemic phase even for $\lambda=1.0$). It means that in a given population where the initial majority supports the vaccination campaign ($D>0.5$), the focus of the government should be to promote a higher engagement of such initial majority. This conclusion corroborates the discussion in \cite{Leask2015}, that indicates that in general the better strategy in a vaccination campaign is not to direct ``battle'' with anti-vaccine individuals. On the other hand, for $D<0.5$ our results suggest that the increase of engagement should not be sufficient to eliminate the disease. In this case, the more efficient strategy is to change the initial conditions, i.e., to convincing more floaters prior to the beginning of the public debate, as discussed by Galam in \cite{Galam2010}. Finally, for $\phi=0.0$ (permanent immunity), the Galam's strategy \cite{Galam2010} could present evident or neglibible results, depending on the initial fraction of pro-vaccine agents and on their engagement.

Summarizing, our results showed that for both analyzed cases (permanent or temporary immunity) the social pressure acts either positively and negatively over the epidemic spreading in our SIS-like model, depending on the initial density of pro-vaccine individuals. Other works \cite{Xia2013,voinson,oraby,Wu2013,Genki2016} also discussed that the social pressure acts as a double-edged sword in vaccination campaigns, but in the mentioned works the authors considered SIR models and they did not take into account the two vaccination schemes considered here (permanent or temporary immunity).

As extensions of this work, one can simulate the model on complex topologies (small world, scale free) in order to analyze the impact of the neighborhood in the dynamics presented here in the fully-connected case. Another interesting extension is to consider more sofisticated dynamics for the opinions' changes, taking into account agents' heterogeneities (conviction, for example \cite{nuno_jstat}) as well as other interesting models as the nonlinear voter model \cite{yang1} and the nonconsensus opinion model \cite{yang2}.


\section*{Acknowledgments}

The authors acknowledge financial support from the Brazilian funding agencies CNPq, CAPES and FAPERJ.


\appendix
\section{}
\label{app}

\qquad The model is definied by the set of ODE's (\ref{eq1}), (\ref{eq2}) and (\ref{eq3}) of the text. In addition, we have the normalization condition $S+V+I=1$.

\subsection{Short time}

\qquad The condition for the epidemic threshold can be obtained from Eq. (\ref{eq2}) taking $t=0$,
\begin{eqnarray}  \nonumber
\frac{dI}{dt} \bigg|_{t=0}  
=
(1-\gamma) \lambda\,S_{o}\,I_{o}\,f_{+1}(0) + \lambda\,S_{o}\,I_{o}\,f_{-1}(0) - \alpha\,I_{o}
\\ \nonumber
=
(1-\gamma)\,\lambda\,S_{o}\,I_{o}\,D + \lambda\,S_{o}\,I_{o}\,(1-D) - \alpha\,I_{o}
\\ \nonumber
=
I_{o}\,\alpha ((1-\gamma\,D)\,S_{o}\,\frac{\lambda}{\alpha} - 1)
\\ \label{app1}
= I_{o}\,\alpha\,(R_{e} -1)
\end{eqnarray}
Comparing the last expression with Eq. (\ref{eq4}), we have $R_{e}=(1-\gamma\,D)S_{o}\,\lambda/\alpha$, that is Eq. (\ref{eq5}) of the text.


\subsection{Long time and $\phi\neq 0 $}

\subsubsection{Case I: $D < 0.5$}
 
\qquad The steady state of the opinion dynamics is $f_{+1}=0$ and $f_{-1}=1$ \cite{Galam1999,galam_mr}, i.e., all agents are Anti-vaccine after a long time. For the epidemic states, the limit $t\to\infty$ for Eqs. (\ref{eq1}), (\ref{eq2}) and (\ref{eq3}) of the text gives us
\begin{eqnarray} \label{app2}
\dot{S_{\infty}} = 0 & = & - \lambda\,S_{\infty}\,I_{\infty}  + \alpha\,I_{\infty}  + \phi\,V_{\infty} ~, \\ \label{app3}
\dot{I_{\infty}} = 0  & = & \lambda\,S_{\infty}\,I_{\infty} - \alpha\,I_{\infty} ~, \\ \label{app4}
\dot{V_{\infty}} = 0  & = & \phi\,V_{\infty} ~.
\end{eqnarray}
Eq. (\ref{app4}) for $\phi\neq 0$ gives us $V_{\infty}=0$. From Eq. (\ref{app3}) we obtain two solutions,
\begin{eqnarray}  \label{app5}
I_{\infty} & = & 0 ~, \\  \label{app6}
S_{\infty} & = & \frac{\alpha}{\lambda} ~.
\end{eqnarray} 
Considering the normalization condition, the solution $V_{\infty}=0$ and the results (\ref{app5}) and (\ref{app6}), one obtains
\begin{align}  \label{app7}
I_{\infty}  =\begin{cases} 0 & \lambda \leq \lambda_{c}
 \\ \frac{1}{\lambda}(\lambda-\lambda_{c}) & \lambda > \lambda_{c} \end{cases}
\end{align}
\noindent
where $\lambda_{c}=\lambda_{c}^{I}=\alpha$ is the epidemic threshold for $D<0.5$, Eq. (\ref{eq11}) of the text.

 
\subsubsection{Case II: $D>0.5$ }

\qquad In this case, the steady state of the opinion dynamics is $f_{+1}=1$ and $f_{-1}=0$ \cite{Galam1999,galam_mr}, i.e., all agents are Pro-vaccine. For the epidemic states, the limit $t\to\infty$ for Eqs. (\ref{eq1}), (\ref{eq2}) and (\ref{eq3}) of the text gives us
\begin{eqnarray} \label{app8}
\dot{S_{\infty}} = 0 & = & - \gamma\,S_{\infty} - (1-\gamma)\,\lambda\,S_{\infty}\,I_{\infty}  + \alpha\,I_{\infty} + \phi\,V_{\infty} ~, \\ \label{app9}
\dot{I_{\infty}} = 0 & = & (1-\gamma)\,\lambda\,S_{\infty}\,I_{\infty} - \alpha\,I_{\infty} ~, \\ \label{app10}
\dot{V_{\infty}} = 0 & = & \gamma\,S_{\infty} - \phi\,V_{\infty} ~.
\end{eqnarray}
\noindent
Eq. (\ref{app9}) gives us
\begin{eqnarray} \label{app11}
I_{\infty} & = & 0 ~, \\ \label{app12}
S_{\infty} &  = & \frac{\alpha}{(1-\gamma)\,\lambda} ~.
\end{eqnarray}
Considering Eqs. (\ref{app10}) and (\ref{app12}), one gets for $\phi\neq 0$
\begin{equation} \label{app13}
V_{\infty} = \frac{\gamma}{\phi}\,\frac{\alpha }{(1-\gamma)\,\lambda} ~.
\end{equation}
Using the normalization condition, one obtains
\begin{equation} \label{app14}
I_{\infty} = \frac{1}{\lambda}\,\Big(\lambda-\alpha\,\frac{\phi + \gamma}{\phi\,(1-\gamma)}\Big) ~.
\end{equation}
Comparing this last result with the standard form $I_{\infty}\sim(\lambda-\lambda_{c})^{\beta}$, one obtains $\lambda_{c}=\lambda_{c}^{II}=\alpha\,\frac{\phi + \gamma}{\phi\,(1-\gamma)}$, that is the epidemic threshold for $D>0.5$, Eq. (\ref{eq12}) of the text, and the typical mean-field exponent $\beta=1$.


\subsection{Long time and $\phi=0$}

\subsubsection{Case I: $D < 0.5$}

\qquad In a similiar way as we made before, the steady-state equations are given by
\begin{eqnarray} \label{app15}
\dot{S_{\infty}} = 0 & = & - \lambda\,S_{\infty}\,I_{\infty}  + \alpha\,I_{\infty} ~, \\ \label{app16}
\dot{I_{\infty}} = 0 & = & \lambda\,S_{\infty}\,I_{\infty} - \alpha\,I_{\infty} ~, \\ \label{app17}
\dot{V_{\infty}} = 0 & = & 0\,V_{\infty} ~.
\end{eqnarray}
\noindent
Eq. (\ref{app17}) is identically null, and in this case we have a set of two equations with three variables, i.e., there is not complete solution. However, from Eq. (\ref{app16}) we have $I_\infty=0$ or $S_\infty=\alpha/\lambda$. The normalization condition gives us the general solution
\begin{align} \label{app18}
I_\infty  =\begin{cases} 0 & \lambda \leq \lambda_{c}
\\ 1 - \frac{\alpha}{\lambda} -V_{\infty} & \lambda > \lambda_{c}
\end{cases}
\end{align} 
\noindent
and one can not obtain an explicity result for the epidemic threshold $\lambda_{c}$, since $V_{\infty}$ can be a function of $\lambda$.


\subsubsection{Case II: $D>0.5$ }

\qquad In a similiar way as we made before, the steady-state equations are given by
\begin{eqnarray} \label{app19}
\dot{S_{\infty}} = 0 & = & - \gamma S_\infty - (1-\gamma) \lambda S_\infty I_\infty  + \alpha I_\infty ~, \\ \label{app20}
\dot{I_{\infty}} = 0 & = & (1-\gamma) \lambda S_\infty I_\infty - \alpha I_\infty ~, \\ \label{app21}
\dot{S_{\infty}} = 0 & = & \gamma S_\infty ~.
\end{eqnarray}
\noindent
For $\gamma\neq 0$ one gets $S_{\infty}=I_{\infty}=0$, and the normalization conditions gives us $V_{\infty}=1$.


\end{document}